\documentclass[twocolumn,prx,superscriptaddress]{revtex4-1}
\usepackage[utf8]{inputenc}
\usepackage{xcolor}
\usepackage{bm}
\usepackage{graphicx}
\usepackage{physics}
\usepackage{amsmath}
\usepackage{amssymb}
\usepackage{times}
\usepackage{enumitem}
\usepackage[colorlinks=true,citecolor=blue]{hyperref}

\newcommand{\jila}{\affiliation{JILA, NIST and Department of Physics, University of Colorado, Boulder, Colorado, USA}}
\newcommand{\ctqm}{\affiliation{Center for Theory of Quantum Matter, University of Colorado, Boulder, Colorado, USA}}

\begin{document}

\title{Time complexity in preparing metrologically useful quantum states}

\author{Carla M. Quispe Flores}
\affiliation{Department of Physics, Colorado School of Mines, Golden, Colorado 80401, USA}

\author{Raphael Kaubruegger}\jila

\author{Minh C. Tran}
\affiliation{IBM Quantum, IBM T.J. Watson Research Center, Yorktown Heights, NY 10598, USA}

\author{Xun Gao} \jila

\author{Ana Maria Rey}\jila\ctqm

\author{Zhexuan Gong}
\email{gong@mines.edu}
\affiliation{Department of Physics, Colorado School of Mines, Golden, Colorado 80401, USA}

\begin{abstract}
We investigate the fundamental time complexity, as constrained by Lieb-Robinson bounds, for preparing entangled states useful in quantum metrology. We relate the minimum time to the Quantum Fisher Information ($F_Q$) for a system of $N$ quantum spins on a $d$-dimensional lattice with $1/r^\alpha$ interactions with $r$ being the distance between two interacting spins. We focus on states with $F_Q \sim N^{1+\gamma}$ where $\gamma \in (0,1]$, i.e., scaling from the standard quantum limit to the Heisenberg limit. For short-range interactions ($\alpha > 2d+1$), we prove the minimum time $t$ scales as $t \gtrsim L^\gamma$, where $L \sim N^{1/d}$. For long-range interactions, we find a hierarchy of possible speedups: $t \gtrsim L^{\gamma(\alpha-2d)}$ for $2d < \alpha < 2d+1$, $t \gtrsim \log L$ for $(2-\gamma)d < \alpha < 2d$, and $t$ may even vanish algebraically in $1/L$ for $\alpha < (2-\gamma)d$. These bounds extend to the minimum circuit depth required for state preparation, assuming two-qubit gate speeds scale as $1/r^\alpha$. We further show that these bounds are saturable, up to sub-polynomial corrections, for all $\alpha$ at the Heisenberg limit ($\gamma=1$) and for $\alpha > (2-\gamma)d$ when $\gamma<1$. Our results establish a benchmark for the time-optimality of protocols that prepare metrologically useful quantum states.

\end{abstract}

\maketitle

\newcommand{\avg}[1]{\langle #1 \rangle} 

\section{Introduction}
Quantum metrology aims to harness quantum entanglement to enhance the precision of measurements beyond classical limits \cite{giovannetti_quantum_2006}. Assuming a system of $N$ quantum spins that act as quantum sensors, each coupling independently to a target signal, the fundamental limit of sensing error scales as $1/\sqrt{N}$ if the sensors are in a classical, unentangled state. This scaling improves to the $1/N$ Heisenberg limit if certain entangled states of the sensors can be prepared \cite{giovannetti_advances_2011}. The limit of measurement precision for any quantum state can be quantified by the quantum Fisher information (QFI), $F_Q$, contained within the state \cite{montenegro2024review}. For metrologically useful entangled states, $F_Q$ scales between $N$ (the standard quantum limit) and $N^2$ (the Heisenberg limit). A prominent example is a spin squeezed state \cite{kitagawa_squeezed_1993}, where $F_Q \sim N^{1+\gamma}$ with $\gamma \in (0,1]$ depending on the type of squeezing \footnote{Throughout this paper, for two positive functions $f(N)$ and $g(N)$ of the system size $N$, the notation $f(N) \sim g(N)$ means that there exist positive constants $c_1$ and $c_2$ and $N_0$ such that for $N \ge N_0$, $c_2 g(N) \le f(N) \le c_1 f(N)$. Similarly, the notation $f(N) \lesssim g(N)$ means that there exists positive constants $c_1$ and $N_0$ such that for $N \ge N_0$, $f(N) \le c_1 f(N)$, while $f(N) \gtrsim g(N)$ means that there exists positive constants $c_2$ and $N_0$ such that for $N \ge N_0$, $f(N) \ge c_2 f(N)$}. Another example is the GHZ state, where $F_Q$ can achieve its maximal possible value of $N^2$.

The main goal of this work is to characterize the complexity of preparing a metrologically useful state with $F_Q \gtrsim N^{1+\gamma}$ and $\gamma>0$ from a classical state. Intuitively, a metrologically useful state must possess non-local correlations and thus cannot be prepared using a finite-depth local quantum circuit or a finite-time evolution of a local Hamiltonian \cite{chen_local_2010}. In practice, metrologically useful states such as spin squeezed states are usually prepared using systems that feature non-local, often all-to-all interactions among particles \cite{sewell_magnetic_2012,bohnet_quantum_2016,luo2024}. However, a rigorous time complexity bound for preparing these states is unknown for general quantum many-body systems. Establishing such a bound is not only important for our fundamental understanding of speed limits in quantum many-body dynamics \cite{anthony_chen_speed_2023}, but also practically useful for identifying optimal experimental protocols in quantum metrology, as faster protocols are less prone to decoherence during state preparation \cite{giovannetti_advances_2011}.

In this work, we identify the minimum time required to prepare a metrologically useful state using any quantum spin Hamiltonian with spin-spin interactions that decay no slower than a power-law $1/r^{\alpha}$, with $r$ denoting the inter-spin distance. Our results include the entire range of $\alpha \in [0,\infty]$, thus incorporating nearly all physical interactions found in current quantum hardware platforms. By establishing a link between the QFI and dynamical two-body correlations during state preparation, we establish rigorous time bounds on preparing a state with a given $F_Q$, utilizing recently developed Lieb-Robinson-type bounds for long-range interacting systems that are optimal for $\alpha>2d$ up to sub-polynomial corrections \cite{tran2021lieb,tran_hierarchy_2020,kuwahara_strictly_2020}. We emphasize that our bounds apply to the preparation of both pure and mixed quantum states, thus allowing for an initial state that is a mixture of product states, which is often the case experimentally.

For different ranges of $\alpha$, we derive rigorous bounds on the scaling of the minimum state preparation time $t$ with the linear system size $L\sim N^{1/d}$ for a given scaling exponent $\gamma$ of $F_Q$. For $\alpha>2d+1$, we find $t \gtrsim L^{\gamma}$ with no scaling dependence on $\alpha$. Reaching the Heisenberg limit ($\gamma=1$) thus requires a preparation time proportional to the linear system size $L$. This is consistent with the existence of linear light cones for the spread of quantum information in such interactions \cite{anthony_chen_speed_2023,kuwahara_strictly_2019,tran_hierarchy_2020}. For $2d<\alpha<2d+1$, we find that a polynomial speedup in state preparation is possible, with $t \gtrsim L^{\gamma(\alpha-2d)}$, while for $(2-\gamma) d<\alpha<2d$, $t \gtrsim \log L$, indicating a possible exponential speedup. For $\alpha<(2-\gamma) d$, the minimum preparation time could vanish polynomially in $1/L$. For example, for $\alpha<d$, we prove that $t \gtrsim 1/L^{(2-\gamma)d-\alpha}$.

\begin{figure}[ht]
\noindent \includegraphics[width=0.47\textwidth]{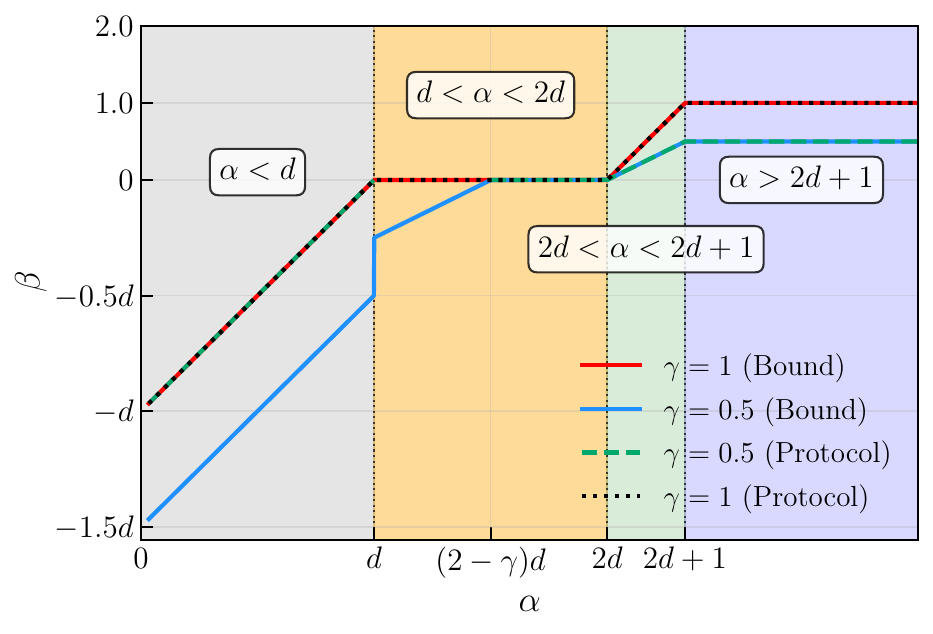}

\caption{Comparison of the time-scaling exponent $\beta$ from our theoretical bound ($t \gtrsim L^{\beta}$) with that of the fastest known protocols ($t \sim L^{\beta}$), ignoring sub-polynomial corrections. The task is to prepare a metrologically useful entangled state ($F_Q \sim N^{1+\gamma}$) using a Hamiltonian with $1/r^{\alpha}$ spin-spin interactions. We compare $\beta_{\text{bound}}$ (solid lines) and $\beta_{\text{protocol}}$ (dashed lines) for both $\gamma=1$ and $\gamma=0.5$. Up to sub-polynomial corrections, our bound is optimal for $\gamma=1$ (where the exponents match) with any $\alpha>0$ or for $0<\gamma<1$ with $\alpha > (2-\gamma)d$.}

\label{fig0}
\end{figure}

We apply our bounds to identify time-optimal physical protocols for preparing metrologically useful states. Specifically, we compare our bounds to the fastest known protocols for preparing GHZ and spin squeezed states, as shown in Fig.\,\ref{fig0}. Up to sub-polynomial corrections in $L$, we find that our bounds can be saturated for all $\alpha\ge0$ and $\gamma=1$ or for $\alpha >(2-\gamma)d$ and $0<\gamma<1$, indicating that our bounds are nearly optimal. For $\alpha >(2-\gamma)d$, we identify a recently developed time-optimal protocol for quantum state transfer and GHZ state preparation \cite{tran_optimal_2021} as the fastest known protocol for creating metrologically useful states with $F_Q \sim N^{1+\gamma}$. For $\alpha<d$ and $\gamma=1$, we find that the two-axis-twisting protocol for preparing a spin squeezed state is the fastest known protocol that nearly saturates our bound, while other commonly used spin squeezing protocols such as one-axis-twisting and twist-and-turn fall short of achieving the speed limits. Our results also show that many recent spin squeezing protocols for short-range or dipolar interactions \cite{Perlin2020,Block2024,koyluoglu2025} appear substantially suboptimal, motivating the search for more time-efficient spin squeezing protocols in these systems.

\section{Main Results}
To introduce our main results, which constrain the minimum time needed to prepare metrologically useful states, we first define the QFI and relate it to two-body correlations in a quantum many-body state. For simplicity, we consider a single-parameter estimation problem where a set of $N$ independent quantum sensors made of quantum spins interact uniformly with an environmental signal, parameterized by $\theta$, that we wish to estimate. Assuming the sensors start in an initial state given by the density operator $\rho_0$, the interaction between the sensors and the signal leads to a state $\rho_{\theta}= e^{-i\theta A}\rho_0 e^{i \theta A}$, where $A\equiv \sum_i A_i$ is known as the generator of the unitary map $\rho_0 \rightarrow \rho_\theta$. Here $A_i$ is an operator that acts only on the $i^{\text{th}}$ spin, with $\lVert A_i \rVert \le O(1)$, assuming a finite interaction strength and time between an individual sensor and the signal ($||\cdot||$ denotes the operator norm). For example, $\theta$ may represent the strength of a surrounding magnetic field in the $z$ direction, and $A_i = \mu \tau S_i^z$ with $\mu$ denoting the magnetic moment of the spin, $S_i^z$ the spin angular momentum of the $i^{\text{th}}$ spin in the $z$ direction, and $\tau$ the interaction time.

The central idea in quantum metrology is to estimate the parameter $\theta$ by preparing and measuring the state $\rho_{\theta}$ $M$ times. Regardless of the particular measurement setting used, the fundamental statistical error in estimating $\theta$ is lower bounded by the so-called quantum Cramer-Rao bound \cite{braunstein_statistical_1994}:
\begin{align} 
\Delta \theta & \ge 1/\sqrt{M F_Q},
\end{align} 
where $F_Q$ denotes the QFI for the initial state $\rho_0$ and the unitary generator $A$. We will ignore the dependence of $\Delta \theta$ on $M$ in the following and focus only on its dependence on the QFI. Assuming an eigenvalue decomposition of $\rho_0 =\sum_{\mu} \lambda_\mu |\lambda_\mu\rangle\langle \lambda_\mu|$, $F_Q$ is given by \cite{montenegro2024review}:
\begin{align} \label{FQ}
F_Q & = 2 \sum_{\mu,\nu,\lambda_\mu+\lambda_\nu \ne 0} \frac{(\lambda_\mu - \lambda_\nu)^2}{\lambda_\mu + \lambda_\nu} \left| \langle \lambda_{\mu} | A | \lambda_\nu \rangle \right|^2 .
\end{align} 

If $\rho_0$ is a classical state, meaning that each eigenstate $|\lambda_\mu\rangle$ of $\rho_0$ with eigenvalue $\lambda_\mu >0$ is a product state of $N$ quantum spins, one can show that the QFI is additive and $F_Q \lesssim N$, yielding the standard quantum limit $\Delta \theta \gtrsim 1/\sqrt{N}$. If $\rho_0$ contains genuine $N$-body entanglement \cite{toth_multipartite_2012}, $F_Q \lesssim N^2$ instead, leading to the Heisenberg limit $\Delta \theta \gtrsim 1/N$. For a general many-body entangled state, we shall assume $F_Q = O(N^{1+\gamma})$, where $\gamma \in [0,1]$, as this is the case for metrologically useful quantum many-spin states explored experimentally \cite{bohnet_quantum_2016,Cox2016,Hosten2016,Block2024,Miller2024,luo2024}.

If $\rho_0$ is a pure state, one can show that $F_Q$ reduces to \cite{montenegro2024review}
\begin{align} 
F_Q & = 4 (\langle A^2 \rangle - \langle A \rangle ^2) . 
\end{align} 

For a mixed state $\rho_0$, we define $C_\mu = \langle \lambda_\mu |A^2 |\lambda_\mu \rangle - \langle \lambda_\mu |A |\lambda_\mu \rangle^2$, which represents the sum of all two-body connected correlations $ C_{ij} =\langle A_i A_j \rangle - \langle A_i\rangle \langle A_j \rangle$ evaluated on the component state $|\lambda_\mu \rangle$. As shown in Appendix \ref{appendix:A} and in Ref.\,\cite{Liu_2014}, $F_Q$ can be bounded by a weighted average of $C_\mu$ as
\begin{align} \label{FvsC}
F_Q & \le  4 \sum_{\mu} \lambda_\mu C_\mu.
\end{align} 

This inequality indicates the QFI is upper bounded by the average total two-body correlations in each component state. We emphasize that the average in Eq.\,\eqref{FvsC} is performed \emph{after} the expectation value, thereby precluding a large QFI using classically correlated states. In other words, if the state $\rho_0$ has a large QFI, then each component pure state must, on average, also have a large QFI.

We start from an arbitrary classical state $\rho(0) = \sum_\mu \lambda_\mu |\lambda_\mu(0)\rangle \langle\lambda_\mu(0)|$ where each $|\lambda_\mu(0)\rangle$ is a product state. We then assume that a possibly time-dependent Hamiltonian or, equivalently, a quantum circuit prepares the target state $\rho_0$ from the classical state $\rho(0)$ in time $t$. Therefore, $\rho_0 = \sum_\mu \lambda_\mu |\lambda_\mu(t)\rangle \langle\lambda_\mu(t)|$. For each state $|\lambda_\mu(t)\rangle$, we will derive upper bounds on $C_\mu(t)$ using the best-known Lieb-Robinson-type (LR) bounds and then establish upper bounds on the state preparation time.

Throughout this paper, we consider the following large family of Hamiltonians:
\begin{align} \label{H}
    H(t^\prime)=\sum_{i,j} h_{ij}(t^\prime), \quad ||h_{ij}(t^\prime)||\le J_0/r^{\alpha},
\end{align}
where $h_{ij}(t^{\prime})$ is an arbitrary operator that acts on spins $i$ and $j$ and may depend on time $t^\prime$, and $r$ denotes the physical distance between spins $i$ and $j$. Most physical Hamiltonians used in current quantum hardware fall into this family \cite{altman_quantum_2021}. For notational simplicity, we will set the maximum interaction strength $J_0=1$ in the following, which implies that time $t$ is measured in units of $1/J_0$.

Let us first assume $\alpha>2d+1$ and apply the best-known LR bound from Ref.\,\cite{tran_hierarchy_2020}, which results in
\begin{align} \label{LR1}
C_{ij}(t) \lesssim \frac{t^{d+1} \log^{2d} r}{r^{\alpha-d}}
\end{align} 
for $t\lesssim r$. In Appendix \ref{appendix:B}, we show that by summing $C_{ij}(t)$ over all pairs $(i,j)$,
\begin{align} \label{C1}
C_\mu(t) \lesssim N t^d
\end{align} 
holds for $t \lesssim L$ and any initial product state $|\lambda_\mu(0)\rangle$. Plugging this inequality into Eq.\,\eqref{FvsC} and using $\sum_\mu \lambda_\mu=1$, we obtain $F_Q \lesssim N t^d$. Since $F_Q \sim N^{1+\gamma}$, we arrive at
\begin{align}  \label{t1}
t \gtrsim L^{\gamma}.
\end{align} 

For $2d<\alpha\le 2d+1$, the best-known LR bound is developed in Ref.\,\cite{tran2021lieb}, which gives
\begin{align} \label{LR2}
C_{ij}(t) \lesssim \left(\frac{t}{r^{\alpha - 2d - \epsilon}} \right)^{\frac{\alpha - d}{\alpha - 2d} - \frac{\varepsilon}{2}} 
\end{align} 
for $t \lesssim r^{\alpha-2d-\epsilon}$, where $\epsilon$ is an arbitrarily small positive constant resulting from sub-polynomial corrections. According to Appendix \ref{appendix:B}, this leads to
\begin{align} \label{C2}
C_\mu(t) \lesssim N t^{\frac{d}{\alpha-2d-\epsilon}}
\end{align} 
and to achieve $F_Q \sim N^{1+\gamma}$, we require
\begin{align} \label{t2}
t \gtrsim L^{\gamma(\alpha-2d-\epsilon)}.
\end{align} 

For $d<\alpha\le 2d$, we first derive a bound on $C_{ij}(t)$ in Appendix \ref{appendix:B} using the best-known LR bound from Ref.\,\cite{hastings_spectral_2006}, which reads
\begin{align} \label{LR3}
C_{ij}(t) \lesssim \frac{e^{vt}-vt-1}{r^{\alpha-d}},
\end{align}
where $v \sim \int_0^L \frac{d^d r}{r^\alpha} $ is proportional to the average interaction energy per spin. For $\alpha>d$, $v\sim 1$. Further derivations in Appendix \ref{appendix:B} show that to achieve $F_Q \sim N^{1+\gamma}$, we require
\begin{align} \label{t3}
t \gtrsim \begin{cases}
    \log L & \alpha> (2-\gamma)d \\
    1 & \alpha = (2-\gamma)d \\
    L^{\frac{\alpha - (2-\gamma)d}{2}} & \alpha < (2-\gamma)d
\end{cases}
\end{align}

For $\alpha \le d$, we can apply the bound 
\begin{align}\label{LR4}
C_{ij}(t) \lesssim vt
\end{align}
derived in Appendix \ref{appendix:B}, where $v \sim \log L$ for $\alpha=d$ and $v\sim L^{d-\alpha}$ for $\alpha<d$. One can show that to achieve $F_Q \sim N^{1+\gamma}$, we require
\begin{align} \label{t4}
t \gtrsim \begin{cases}
    L^{d(\gamma-1)}/\log L & \alpha=d \\
    L^{\alpha - (2-\gamma)d} & \alpha < d
\end{cases}
\end{align}

\section{Discussion}

A key question is whether the time complexity bounds derived in Section II are tight or optimal, meaning that they can be saturated, up to a constant factor, by explicit Hamiltonians or quantum circuits. We can also relax the bounds' optimality condition by allowing sub-polynomial (e.g., logarithmic) corrections in $N$ or $L$, as adopted in the study of optimal long-range Lieb-Robinson bounds \cite{tran_hierarchy_2020,tran2021lieb,tran_optimal_2021}. In this sense, we will first show that the time complexity bounds derived in Section II are optimal for $\gamma=1$. We have thus obtained (near) optimal time complexity bounds for preparing metrologically useful states that achieve the Heisenberg limit. Explicitly, for $\gamma=1$ we have
\begin{align} \label{HLbounds}
t \gtrsim \begin{cases}
   L & \alpha>2d+1 \\
   L^{\alpha-2d-\epsilon} & 2d<\alpha\le 2d+1 \\
    \log L & d< \alpha \le 2d \\
    1 /\log L & \alpha = d \\
    L^{\alpha - d} & 0\le \alpha < d
\end{cases}
\end{align} 

To saturate the above time complexity bounds, we employ two different protocols. For $\alpha>d$, we use the protocol from Ref.\,\cite{tran_optimal_2021} that prepares an $N$-qubit GHZ state recursively in a large hypercube from the GHZ-like states of smaller hypercubes. The preparation time scales as:
\begin{align} \label{GHZ}
t_{\text{GHZ}} \sim \begin{cases}
   L & \alpha>2d+1 \\
   L^{\alpha-2d} & 2d<\alpha < 2d+1 \\
   e^{3\sqrt{d\log L}} &\alpha=2d \\
    (\log L)^{\kappa_{\alpha}} & d<\alpha <2d \\
\end{cases}
\end{align} 
where $\kappa_{\alpha} = \log4/\log(2d/\alpha) \in (2,\infty)$ for $\alpha \in (d,2d)$. We see that the bounds in Eq.\,\eqref{HLbounds} are saturated up to sub-polynomial corrections in $L$ for $\alpha>d$.

For $\alpha \le d$, we use a two-axis-twisting (TAT) Hamiltonian to prepare an optimal spin squeezed state with $F_Q \sim N^2$ \cite{kitagawa_squeezed_1993}. Explicitly, we use the following Hamiltonian
\begin{align} \label{TAT}
    H_{\text{TAT}} = J_\alpha \sum_{i,j} (\sigma_i^y \sigma_j^z + \sigma_i^z \sigma_j^y),
\end{align}
where $J_\alpha = L^{-\alpha}$ so that $H_{\text{TAT}}$ belongs to the family of Hamiltonians defined in Eq.\,\eqref{H}. Starting from an initial state polarized in the $x$ direction, it has been shown that at time $J_\alpha t\sim \log(N)/N$, $F_Q \sim N^2$ and the Heisenberg limit is reached \cite{kajtoch_quantum_2015}. The preparation time is thus
\begin{align} \label{squeezed}
t_{\text{squeezed}} \sim  \log (L) L^{\alpha-d} ,
\end{align} 
which saturates our bounds in Eq.\,\eqref{HLbounds} up to sub-polynomial corrections in $L$ for $\alpha\in [0,d]$. Note that using a true $1/r^{\alpha}$ interaction in Eq.\,\eqref{TAT}, which appears naturally in experimental platforms such as trapped ions \cite{montenegro2024review}, is unlikely to speed up this protocol, as it would break the permutation symmetry required to prepare highly spin squeezed states \cite{Block2024,foss-feig_entanglement_2016}. We also note that an approximate GHZ state may be prepared with the same time complexity using protocols developed recently \cite{yin_fast_2025,ma_quantum_2025,zhang_fast_2024} which are based on TAT. These protocols can also be used to saturate our bounds in a similar fashion.

Next, for $0<\gamma<1$, we can show that our bounds are optimal for $\alpha >(2-\gamma)d$ up to sub-polynomial corrections. Here, instead of preparing the full $N$-qubit system into a GHZ state, we divide the system into $N^{1-\gamma}$ blocks of $N^{\gamma}$ qubits each. For sufficiently large $N$, this division is always possible with negligible rounding errors. For each block, we apply the same time-optimal GHZ state preparation protocol from Ref.\,\cite{tran_optimal_2021} for $\alpha>(2-\gamma)d$, which requires a preparation time given by Eq.\,\eqref{GHZ} upon replacing $L$ with $L^{\gamma}$. The final state of the system is a tensor product of $N^{1-\gamma}$ GHZ states, each containing $N^{\gamma}$ qubits. Therefore $F_Q \sim N^{1+\gamma}$, and the state preparation time is given explicitly by:
\begin{align} \label{GHZ2}
t_{\text{GHZ}}(\gamma) \sim \begin{cases}
   L^{\gamma} & \alpha>2d+1 \\
   L^{\gamma(\alpha-2d)} & 2d<\alpha < 2d+1 \\
   e^{3\sqrt{\gamma d\log L}} &\alpha=2d \\
    (\gamma \log L)^{\kappa_{\alpha}} & (2-\gamma) d<\alpha <2d \\
\end{cases}
\end{align} 
This protocol saturates our bounds in Section II for any $0<\gamma<1$ up to sub-polynomial corrections for $\alpha>(2-\gamma)d$. However, for $d<\alpha\le (2-\gamma)d$, this protocol still gives a state preparation time that increases with $L$, which does not saturate the bounds in Eq.\,\eqref{t3} that decrease with increasing $L$.

For $\alpha\le d$, while the TAT protocol nearly saturates our bounds for $\gamma=1$, it cannot be modified to saturate our bounds for $0<\gamma<1$ by dividing the system into blocks. This is because the TAT protocol takes \emph{longer} to maximally squeeze a smaller number of qubits for $\alpha \le d$, so the preparation time on each block already exceeds the bound in Eq.\,\eqref{t4}. Alternatively, one can perform the TAT protocol on the entire system but stop at an earlier time to obtain a state with $F_Q \sim N^{1+\gamma}$. An approximate analytical calculation of $F_Q$ based on $H_{\text{TAT}}$ for $\alpha=0$ and $t\ll 1/\sqrt{N}$ gives \cite{kajtoch_quantum_2015}:
\begin{align} \label{FQt}
F_Q(t) \approx N e^{2Nt-\frac{\sinh(2Nt)}{N}+\frac{t}{\sqrt{N}}}.
\end{align} 
This leads to $F_Q \sim N^{1+\gamma}$ for $t \sim \gamma \log(N)/N$, or equivalently, $t \sim \gamma \log(L) L^{\alpha-d}$ for $\alpha>0$. Less time is indeed required for preparing a state with lower QFI with this protocol. However, it still fails to saturate the bound in Eq.\,\eqref{t4}, which predicts an $N^{\gamma}$ dependence on the preparation time instead of a linear dependence on $\gamma$.

We have also examined other widely used spin squeezing protocols, such as one-axis-twisting (OAT) \cite{kitagawa_squeezed_1993} and twist-and-turn (TnT) \cite{julia2012dynamic}, which can create metrologically useful states with $\gamma=2/3$ and $\gamma=1/2$, respectively. However, these protocols do not saturate the bound in Eq.\,\eqref{t4} either. We provide a detailed comparison of the optimal state preparation time and $F_Q$ scaling for the three common spin squeezing protocols (TAT, OAT, and TnT) in Appendix \ref{appendix:C}, based on exact numerical calculations for hundreds of spins.

\section{Conclusion and Outlook}

In this paper, we derived rigorous lower bounds on the time complexity for preparing metrologically useful quantum states using any two-body interacting quantum spin Hamiltonian with short- or long-range interactions. Our bounds establish a link between the minimum state preparation time and the amount of QFI contained in the state. For a large class of Hamiltonians and metrologically useful states, our bounds are nearly optimal. 

Specifically, for systems with strongly long-range interactions (i.e., $\alpha \le d$), our results indicate that TAT or its variants \cite{yin_fast_2025,zhang_fast_2024} are likely the time-optimal protocols for creating metrologically useful states. Recent advances in Floquet engineering have led to the experimental demonstration of the TAT Hamiltonian with cold atoms or molecules \cite{Miller2024,luo2024}, paving the way towards the optimal generation of metrologically useful quantum states in these systems. 

For systems with shorter-range interactions, especially when $\alpha >2d$, our results suggest that the optimal GHZ state preparation protocol in Ref.\,\cite{tran_optimal_2021} is likely the fastest method for creating metrologically useful quantum states. For the preparation of states with $0<\gamma<1$ and with $\alpha<(2-\gamma)d$, we suspect our bounds are not tight, although improving the bounds remain a challenge due to the lack of optimal Lieb-Robinson-type bounds in this regime \cite{andrew_y_guo_signaling_2020,tran2021lieb}. In addition, many recent spin squeezing protocols \cite{Perlin2020,Block2024,koyluoglu2025} for short-range interacting systems or dipolar interactions ($\alpha=3$) are far from saturating our bounds. Since spin squeezed states are much more robust than GHZ states in metrology applications, it remains an important open question whether time-optimal spin squeezing protocols exist for such relatively short-ranged interactions. A common assumption is that a significant slowdown of spin squeezing is expected for a large $\alpha$ due to the breakdown of permutational symmetry. However, this may not be true if one can adapt the time-optimal GHZ state preparation protocol in Ref.\,\cite{tran_optimal_2021} to prepare certain spin squeezed states with a similar time scale, as recent works have shown that GHZ state preparation is highly related to maximal spin squeezing \cite{yin_fast_2025,ma_quantum_2025,zhang_fast_2024}. On the other hands, it is also possible that our bounds may not be saturable for preparing spin squeezed states when $\alpha>d$, and a meaningful future direction is to derive tighter and saturable time complexity bounds for the preparation of specific classes of metrologically useful states with physically relevant Hamiltonians.

Another major open question is whether we can close the sub-polynomial gaps between our bounds and the fastest known protocols in the regimes where our bounds are nearly optimal. Such gaps also exist between long-range LR bounds and the fastest known quantum state transfer protocols for $\alpha>d$. For $\alpha<d$, the gap was previously expected to be much larger, as the best-known long-range LR bound \cite{andrew_y_guo_signaling_2020} predicts a quantum state transfer time of $t\gtrsim \log(N)/N^{1-\alpha/d}$, while the best-known protocol has $t \sim O(1)$. Our work, in fact, also shows rigorously that the LR bound in Ref.\,\cite{andrew_y_guo_signaling_2020} is optimal up to a $\log(N)$ factor, since the preparation of the optimally squeezed state using Eq.\,\eqref{TAT} generates $C_{ij}\sim 1$ for any two spins $i$ and $j$ in time $t\sim \log(N)/N^{1-\alpha/d}$, while the LR bound in Ref.\,\cite{andrew_y_guo_signaling_2020} implies our Eq.\,\eqref{LR4}, which further leads to $t\sim 1/N^{1-\alpha/d}$. Closing this sub-polynomial gap may be possible if a better method to bound $C_{ij}$ (or $\sum_{ij} C_{ij}$) for $\alpha<d$ can be developed. This may also be useful in proving the well-known fast scrambling conjecture \cite{sekino_fast_2008}, where a $\log(N)$ gap also exists between the best bound \cite{yin_bound_2020} and the fastest known scrambling protocol \cite{lashkari_towards_2013}.

\vspace{-10pt}
\begin{acknowledgments}
We acknowledge funding support from AFOSR FA9550-24-1-0179, ARO W911NF24-1-0128, NSF PHY-2112893, PHY-2317149, and the W. M. Keck Foundation.
\end{acknowledgments}
\bibliographystyle{apsrev4-1}
\bibliography{main}

\appendix \onecolumngrid

\section{Proof of Eq.\,\eqref{FvsC}} \label{appendix:A}

To prove Eq.\,\eqref{FvsC}, we first separate the right-hand side (r.h.s.) of Eq.\,\eqref{FQ} into contributions where both $\lambda_\mu$ and $\lambda_\nu$ are nonzero and where one of them is zero.
\begin{align} \label{F_2}
F = 2 \sum_{\mu,\nu; \lambda_\mu, \lambda_\nu > 0}
\frac{(\lambda_\mu - \lambda_\nu)^2}{\lambda_\mu + \lambda_\nu}
\left|\langle \lambda_\mu | A | \lambda_\nu \rangle\right|^2 + 4 \sum_{\mu,\lambda_\mu > 0 } \lambda_\mu \sum_{\nu,\lambda_\nu=0} \left| \langle \lambda_\mu | A | \lambda_\nu \rangle \right|^2.
\end{align}
where we used the symmetry in Eq.\,\eqref{FQ} with respect to exchanging $\mu$ and $\nu$. We now note that the second sum can be rewritten using the projector onto the null space of $\rho_0$ as
\begin{align}
 & 4 \sum_{\mu,\lambda_\mu > 0 } \lambda_\mu \langle \lambda_\mu |  A ( \mathbb{I} - \sum_{\nu,\lambda_\nu > 0} |\lambda_\nu\rangle\langle\lambda_\nu| ) A | \lambda_\mu \rangle \\
= \, & 4 \sum_{\mu,\lambda_\mu > 0 } \lambda_\mu \left( \langle \lambda_\mu | A^2 | \lambda_\mu \rangle - \langle \lambda_\mu | A | \lambda_\mu \rangle ^2 - \sum_{\nu,\lambda_\nu > 0, \nu\ne \mu} |\langle \lambda_\mu | A | \lambda_\nu \rangle|^2 \right )
\end{align}
Plugging this into Eq.\,\eqref{F_2} and defining $C_\mu =  \langle \lambda_\mu | A^2 | \lambda_\mu \rangle - \langle \lambda_\mu | A | \lambda_\mu \rangle^2$, we have
\begin{align}
F = 4 \sum_{\mu} \lambda_\mu C_\mu - 2 \sum_{\mu,\nu; \mu \ne \nu, \lambda_\mu, \lambda_\nu > 0} \zeta_{\mu\nu} |\langle \lambda_\mu | A | \lambda_\nu \rangle|^2 
\end{align}
where the coefficient $\zeta_{\mu\nu}$ is strictly non-negative 
\begin{align}
\zeta_{\mu\nu} \equiv \lambda_\mu + \lambda_\nu - \frac{(\lambda_\mu - \lambda_\nu)^2}{\lambda_\mu + \lambda_\nu} \ge 0
\end{align}
As a result, Eq.\,\eqref{FvsC} holds.

\section{Proof of time complexity bounds} \label{appendix:B}

In this appendix, we prove the time complexity bounds from Section II for preparing a state with $F_Q\sim N^{1+\gamma}$ using the best-known LR bounds for different ranges of $\alpha$. A key method is to separate the sum in $C_\mu (t) =\sum_{i,j} C_{ij}(t)$ into two parts for a given $t$: one with $r$ (the distance between $i$ and $j$) outside the light cone, and the other with $r$ inside the light cone. The LR bounds usually apply only to the region outside the light cone; inside the light cone, we apply the trivial bound $C_{ij}(t) \lesssim 1$. Since LR bounds are typically translationally invariant, we can also reduce the sum over both $i$ and $j$ to a single sum over $r=0,1,\cdots, L-1$ where $L\sim N^{1/d}$, i.e.
\begin{align}
    C_\mu (t) =\sum_{i,j}C_{ij}(t) \lesssim N \sum_r C_r(t)
\end{align}

Specifically, let us first consider the $\alpha>2d+1$ case. The best-known LR bound in this case is given by Eq.\,\eqref{LR1}. Here the light cone is defined by $t\lesssim r$, so we can perform the following separation:
\begin{align} \label{C1s}
\sum_{i,j} C_{ij}(t) & = N \sum_{r} C_r(t) = N \left(\sum_{r \gtrsim t} C_{ij}(t) + \sum_{r \lesssim t} C_{ij}(t) \right)
\end{align}
As mentioned, the second sum can be trivially bounded by $O(t^d)$. For the first sum, applying Eq.\,\eqref{LR1} leads to
\begin{align}
\sum_{r \gtrsim t} C_{ij}(t) \lesssim \int_t ^L \frac{t^{d+1} \log^{2d} r}{r^{\alpha-d}} d^d r \sim t^{d+1} / L^{\alpha-2d-\epsilon}
\end{align}
where $\epsilon$ is an arbitrarily small positive number to account for sub-polynomial corrections. 

Now we note that for the above sum to exist, we can only consider $t\lesssim L$. However, with $t\lesssim L$ and $\alpha>2d+1$, the second sum in Eq.\,\eqref{C1s} always dominates over the first, thus we end up with Eq.\,\eqref{C1} of the main text.

For $2d<\alpha<2d+1$, the light cone is instead at $t \sim r^{\alpha-2d-\epsilon}$ based on the LR bound in Eq.\,\eqref{LR2}, and we similarly divide $\sum_r C_r(t)$ into two parts, with each part bounded by
\begin{align} \label{C2s}
\sum_{r \gtrsim t^{1/(\alpha-2d-\epsilon)}} C_{ij}(t) & \lesssim \int_0^L \left(\frac{t}{r^{\alpha - 2d-\epsilon}} \right)^{\frac{\alpha - d}{\alpha - 2d}-\frac{\epsilon}{2}}  d^d r
\sim t^{\frac{\alpha - d}{\alpha - 2d}-\frac{\epsilon}{2}} / L^{\alpha-2d-\epsilon^\prime}  \\
\sum_{r \lesssim t^{1/(\alpha-2d-\epsilon)}} C_{ij}(t) & \lesssim t^{d/(\alpha-2d-\epsilon)}
\end{align}
where $\epsilon^\prime$ is another arbitrarily small positive number. Again, we only need to consider $t \lesssim L^{\alpha-2d-\epsilon}$ as otherwise the first sum above does not exist. It is easy to see that for $2d<\alpha<2d+1$, the second sum above also dominates over the first, and we end up with Eq.\,\eqref{C2}.

Next, for $0<\alpha<2d$, the speed of quantum information could diverge exponentially with distance \cite{hastings_spectral_2006,tran_optimal_2021}, making the separation of $\sum_r C_r(t)$ based on the light cone edge ineffective. In addition, the known LR bounds in this regime \cite{hastings_spectral_2006,gong_persistence_2014,andrew_y_guo_signaling_2020} do not directly bound the two-point correlation $C_{ij}(t)$. Instead, they bound the unequal-time commutator $||[A_i(t),A_j||$ for two arbitrary operators $A_i$ and $A_j$ supported on sites $i$ and $j$, respectively, at $t=0$:
\begin{align} \label{AtB}
    ||[A_i(t),A_j]|| \lesssim \frac{e^{vt}-1}{r^{\alpha}}
\end{align}
where $r$ again denotes the distance between spins $i$ and $j$, and $v \sim \max_i \sum_j ||h_{ij}|| \sim \int_0^L \frac{d^d r}{r^\alpha}$. 

To convert this LR bound to a bound on $C_{ij}(t)$, we first define an operator $\tilde{A}_i(t)=e^{iH_{i}t}Ae^{-iH_{i}t}$ which represents the operator $A_i$ evolved under the Hamiltonian $H_{i}\equiv \sum_{k,l; r_{ik},r_{il}<r/2 }h_{kl}$ that contains the interactions from the original Hamiltonian restricted to all spins within a distance $r/2$ from spin $i$. We similarly define $\tilde{A}_j(t)=e^{iH_{j}t}A_j e^{-iH_{j}t}$. Since the supports of $\tilde{A}_i(t)$ and $\tilde{A}_j(t)$ are non-overlapping, one can show that
\begin{equation}
\label{eq:cexpr}
C_{ij}(t) \leq 2( \Vert A_i(t)-\tilde{A}_i(t)\Vert\Vert A_j\Vert+\Vert A_j(t)-\tilde{A}_j(t)\Vert\Vert A_i\Vert).
\end{equation}
We then bound $\Vert A_i(t)-\tilde{A}_i(t)\Vert$ using the triangle inequality as:
\begin{eqnarray} \label{Adiff}
\Vert A_i(t)-\tilde{A}_i(t)\Vert & \le & \int_{0}^{t}d t^{\prime} \sum_{k,l; r_{ik}<r,r_{il}\ge r} \Vert[\tilde{A}_i(t^{\prime}),h_{kl}]\Vert
\end{eqnarray}
Each summand above can be bounded using the LR bound in Eq.\,\eqref{AtB} as
\begin{equation}
\lVert [\tilde{A}_i(t),h_{kl}]\rVert \lesssim \frac{e^{vt}-1}{r_{kl}^\alpha r_{ik}^{\alpha}}
\label{eq:S11}
\end{equation}
Next, we invoke a so-called reproducibility condition, first introduced in Ref.\,\cite{hastings_spectral_2006}, which shows that for any $\alpha > d$
\begin{align}
    \sum_{k} \frac{1}{r_{kl}^\alpha r_{ik}^{\alpha}} \lesssim \frac{1}{r_{il}^\alpha}
\end{align}
Using this in Eq.\,\eqref{Adiff} leads to
\begin{align}
\Vert A_i(t)-\tilde{A}_i(t)\Vert & \lesssim  \int_{0}^{t}d t^{\prime} \sum_{l,r_{il}\ge r} \frac{e^{vt^{\prime}}-1}{r_{il}^\alpha} \lesssim \frac{e^{vt}-1-vt}{r^{\alpha-d}}
\end{align}  
Combined with Eq.\,\ref{eq:cexpr} and a similar bound for $\Vert A_j(t)-\tilde{A}_j(t)\Vert$, we prove Eq.\,\eqref{LR3}, which further leads to:
\begin{align} \label{C3r}
\sum_r C_r(t) & \lesssim (e^{vt}-1-vt) L^{2d-\alpha}
\end{align}

Setting $\sum_r C_r(t) \sim N^\gamma$ in Eq.\,\eqref{C3r} leads to $e^{vt}-vt -1\gtrsim L^{\alpha -(2-\gamma)d}$. If $\alpha > (2-\gamma)d$, we simply have $t \gtrsim \log L$. If $\alpha=(2-\gamma)d$, we have $t \gtrsim 1$, while if $\alpha<(2-\gamma)d$, we have $(vt)^2 \gtrsim L^{\alpha -(2-\gamma)d} \rightarrow 0$ in the large $L$ limit. This proves Eq.\,\eqref{t3} of the main text. 

For $\alpha \le d$, we instead derive a simple bound on $C_{ij}(t)$ independent of $r$ as follows. First, we note that $C_{ij}(t) \lesssim \max_{A_i,B}||[A_i(t),B]||$ where $A_i$ is an arbitrary unity-norm operator on site $i$ and $B$ is an arbitrary unity-norm operator supported on all other sites. This is a very crude bound that essentially bounds how much the support of any operator on site $i$ can spread to other sites. It is not hard to show that \cite{hastings_locality_2010}
\begin{align} \label{C4}
   ||[A_i(t),B]|| \lesssim \int_0^t \sum_{j, j\ne i}||h_{ij}(t^\prime)||dt^\prime \lesssim v t
\end{align}
where $v\sim \log L$ for $\alpha=d$ and $v\sim L^{d-\alpha}$ for $\alpha<d$. As a result, we have $\sum_r C_r(t) \lesssim Nvt$, and setting $\sum_r C_r(t) \sim N^\gamma$ leads to $t \sim L^{d(\gamma-1)}/v$, which proves Eq.\,\eqref{t4}.

\section{Comparison between the bounds and common spin squeezing protocols} \label{appendix:C}

In this appendix, we connect our time-complexity bounds for preparing metrologically useful states to several widely used spin squeezing protocols. We also perform exact numerical calculations for up to $N=1000$ spin-1/2 particles to explicitly compare the bounds to the protocols. Specifically, we consider the following three spin squeezing protocols: (a) two-axis twisting (TAT) \cite{kajtoch2015quantum}, (b) twist-and-turn (TnT) \cite{julia2012dynamic}, and (c) one-axis twisting (OAT) \cite{kitagawa_squeezed_1993}. Since these protocols are typically applied to systems with all-to-all spin interactions (i.e., $\alpha=0$), we use collective spin operators for a system of $N$ spin-1/2 particles, $S_{\alpha} = \sum_{i=1}^N S_i^{\alpha}$ where $\alpha=x,y,z$. The Hamiltonians for these three protocols are respectively given by:
\begin{align}
\hat{H}_{\mathrm{TAT}} &= \chi\,(\hat{S}_y \hat{S}_z + \hat{S}_z \hat{S}_y),\\
\hat{H}_{\mathrm{TnT}} &= \chi\,\hat S_z^{2} - B \hat S_x,\\
\hat{H}_{\mathrm{OAT}} &= \chi\, \hat{S}_z^2,
\end{align}
where $\chi$ denotes the interaction strength, which can be set to unity to match Eq.\,\eqref{H} used for our bounds.

For all protocols, we assume an initial state where all spins are polarized in the $x-$direction. Changing the polarization to the $+x$ direction would not affect the results for the TAT and OAT Hamiltonians, but scalable squeezing could no longer be generated using the TnT Hamiltonian for $B>0$. The goal of these squeezing protocols is to minimize the variance of the collective spin angular momentum operator in some direction of the $y-z$ plane (transverse to the initial spin direction). Orthogonal to this direction in the same plane, we expect the collective spin angular momentum operator, $S_{\theta} = S_{y}\cos\theta + S_{z}\sin\theta$, to have the maximal variance instead. Due to the $\mathbb{Z}_2$ symmetry of the initial state and the three Hamiltonians above upon flipping the $y$ and $z$ components of the collective spin, we have $\langle S_{y}\rangle=\langle S_{z}\rangle=\langle S_{\theta}\rangle=0$ at all times. Ignoring any coupling to the environment, the state of the system remains pure and the QFI is simply given by $F_Q = 4\langle S_{\theta}^2 \rangle$ when the probe Hamiltonian is proportional to $S_{\theta}$.

The angle $\theta$ that maximizes the QFI (which also maximizes the amount of squeezing) depends on the evolution time $t$ and the particular Hamiltonian used. To obtain $\theta$, we numerically compute $\langle S_{y}^{2}\rangle$, $\langle S_{z}^{2}\rangle$, $\langle S_{y} S_{z}\rangle$, and $\langle S_{z} S_{y}\rangle$ of the state at time $t$ for each Hamiltonian, and maximize $F_Q$ over $\theta$ according to
\begin{equation}
F_Q(\theta)= 4\left [
\cos^{2}\theta\,\langle S_{y}^{2}\rangle
+\sin^{2}\theta\,\langle S_{z}^{2}\rangle
+\sin\theta\cos\theta\,\langle S_{y}S_{z}+S_{z}S_{y}\rangle\right ]
\end{equation}

This maximization of $F_Q$ is performed for each time point $t$ in our numerical calculation. We then further maximize $F_Q$ by scanning over the evolution time $t$ to find the optimal time numerically. This process is facilitated by existing knowledge of the optimal spin-squeezing time $t_{\text{opt}}$ \cite{kajtoch2015quantum,kitagawa_squeezed_1993,julia2012dynamic}. Roughly speaking, $\chi t_{\text{opt}} \sim \ln(N)/N$ for both the TAT and TnT protocols, and $\chi t_{\text{opt}} \sim N^{-2/3} $ for the OAT protocol. For the TnT protocol, we set $B=2\chi N$, which achieves the maximal scaling of squeezing in $N$ \cite{kajtoch_quantum_2015}.

\begin{figure}[ht]
\centering
\includegraphics[width=0.49\textwidth]{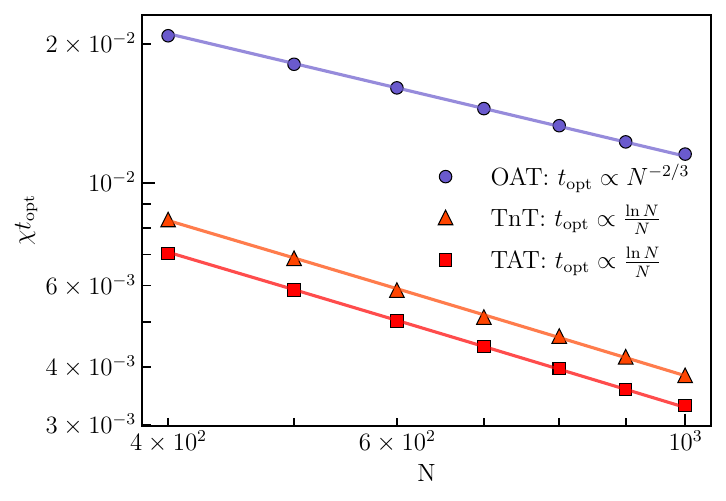}
\hfill
\includegraphics[width=0.49\textwidth]{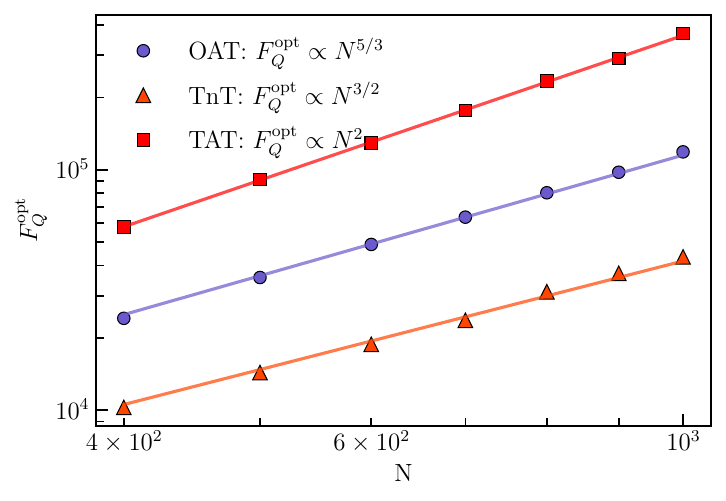}
\caption{Optimal spin-squeezing time $t_{\text{opt}}$ (left) and the corresponding QFI $F_{Q}^{\text{opt}}$ (right) as a function of the number of spins $N$ for three spin squeezing protocols: (a) two-axis twisting (TAT), (b) twist-and-turn (TnT), and (c) one-axis twisting (OAT). The data points are fitted using the following models and the fitted models are shown as solid lines: $\chi t_{\text{opt}} = A N^{-2/3}$ with $A = 1.144 \pm 0.003$ for OAT, $\chi t_{\text{opt}} = A \ln(N)/N$ for TAT and TnT, with $A = 0.4730 \pm 0.0009$ and $A = 0.554 \pm 0.001$ respectively; $F_{Q}^{\text{opt}} = A N^2$ with $A = 0.3627 \pm 0.0014$ for TAT, $F_{Q}^{\text{opt}} = A N^{3/2}$ with $A = 1.32 \pm 0.02$ for TnT, and $F_{Q}^{\text{opt}} = A N^{5/3}$ with $A = 1.152 \pm 0.009$ for OAT.} \label{fig1}
\end{figure}

For $N=400$ to $N=1000$ spins, we numerically compute $t_{\text{opt}}$ and the corresponding QFI $F_{Q}^{\text{opt}}$. The scaling of $t_{\text{opt}}$ and $F_{Q}^{\text{opt}}$ with $N$ is shown in Fig.\,\ref{fig1}, which agrees with previous analytical calculations \cite{kajtoch2015quantum,kitagawa_squeezed_1993,julia2012dynamic}. From these figures, it is clear that the TAT protocol performs best: it not only requires the least state preparation time but also creates a state with the highest QFI or metrological usefulness. Importantly, it saturates our time-complexity bounds $t \gtrsim 1/N$ for $\alpha=0$ and $\gamma=1$ up to a logarithmic factor in $N$. The TnT and OAT protocols generate states with $\gamma=1/2$ and $\gamma=2/3$, respectively, but they do not saturate our time complexity bounds, which predict $t_{\text{opt}} \gtrsim 1/N^{2-\gamma}$. However, it should also be noted out that the TAT protocol is more challenging to implement experimentally compared to the TnT and OAT protocols \cite{julia2012dynamic,Miller2024,luo2024}.

\end{document}